\documentclass[superscriptaddress,twocolumn,showpacs,prb,floatfix]{revtex4}
\usepackage{epsfig}
\begin{document}

\title{Ordering temperatures of Ising Spin Glasses}
\author{I.~A.~Campbell}
\affiliation{Laboratoire des Verres, Universit\'e Montpellier II,
34095 Montpellier, France}
\date{\today}

\begin{abstract}
Exploiting an approach due to Singh and Fisher I show that in the high dimension limit the ordering temperature of near neighbour Ising Spin Glasses drops linearly with the kurtosis of the interaction distribution, in excellent agreement with accurate high temperature series data of Daboul, Chang and Aharony. At lower dimensions the linear relation no longer applies strictly but the kurtosis can still be taken to be an appropriate parameter for ranking different systems. I also compare the series estimates with simulation and Migdal-Kadanoff estimates where these are available. 
\end{abstract}

\maketitle


In order to establish precise values of critical exponents at a continuous transition, it is essential to first obtain accurate and reliable values for the critical temperature $T_c$. 
The ordering temperatures $T_c$ of Ising spin glasses (ISGs) are not expected to be universal but to depend on 
the form of the interaction distribution. In ISGs with random near neighbour interactions of zero mean $\langle J_{ij} \rangle = 0$ the interaction distributions are normalized according to the standard convention $\langle J_{ij}^2 \rangle = 1$ (where $ \langle .. \rangle $ represents an average over the interaction distribution $J_{ij}$ between pairs of spins). It was suggested on purely empirical grounds \cite{bernardi:97}  that $T_c$ values should be ranked from highest to lowest according to the value of the kurtosis of the interaction distribution, 
\begin{equation}
K = \langle J_{ij}^4 \rangle /\langle J_{ij}^2 \rangle ^2
\label{kurtosis}
\end{equation}
(An alternative definition of kurtosis, sometimes referred to as the excess kurtosis, corresponds to $K - 3$ which leads to a zero value for the Gaussian distribution. For present purposes it turns out that the the definition in Eq.~(\ref{kurtosis}) is more convenient.)
Extensive high temperature series calculations by Daboul, Chang and Aharony provide 
accurate values of $T_c$ in dimensions $d$ from $8$ to $4$ which are consistent with this phenomenological 
rule in each dimension \cite{daboul:04}. 

Exploiting an approach due to Singh and Fisher \cite{singh:88} I show rigorously that in the high dimension limit the ordering temperature of near neighbour ISGs drops linearly with the kurtosis of the interaction distribution, and find excellent quantitative agreement between the Singh-Fisher predictions and the data of Daboul {\it et al} \cite{daboul:04} in dimensions $7$ and $8$. At lower dimensions the linear relation no longer applies strictly but the trend remains and the kurtosis can still be taken to be an appropriate parameter for ranking different systems. 

Singh and Fisher \cite{singh:88} used a $1/d$ expansion technique (whose principle largely pre-dates RGT \cite{fisher:64} ) to evaluate $T_c$ in ISGs as a function of dimension and of interaction distribution. On this approach and for the near neighbour ISG with bimodal $\pm J$ interactions, they define 
\begin{equation}
\omega_c(\pm J) = \tanh^2[J/T_c(\pm J)]
\label{omegaJ}
\end{equation}
and find that
\begin{equation}
[\omega_c]^{-0.5} = (\sigma)^{0.5}[1 - (7/2)/\sigma^2 - (21/2)/\sigma^3 - ...]
\end{equation}
where $\sigma = (2d-1)$.
They summed this series, truncating appropriately, to estimate the sequence of $T_c(\pm J)$ values for the bimodal $\pm J$ interaction in different dimensions. They obtained satisfactory agreement with the high temperature series values which had just been been evaluated for this interaction \cite{singh:86,singh:87,singh:87b}. More recent high temperature series work \cite{klein:91,daboul:04} gives values consistent with the earlier series estimates for the bimodal distribution.   

For present purposes the important step of Singh and Fisher was the next one : they state that for an arbitrary interaction distribution that we will label $X$, the mean over the interaction distribution $J_{ij}(X)$ of
\begin{equation}
\omega_c = \langle \tanh^{2}[J_{ij}(X)/T_c(X)] \rangle
\label{omegaX}
\end{equation} 
is the same for all distributions to order $1/\sigma^{2}$. This is an implicit equation for $T_c(X)$ once $\omega_c$ is known. The leading deviation was given as  
\begin{equation}
[\omega_c(X)-\omega_c(\pm J)] = \omega_c(\pm J) [1 - 3(\mu_2 -1)/\sigma^3) ...]
\label{deviation}
\end{equation}
where 
\begin{equation}
\mu_n = \langle \tanh^{2n}[J_{ij}/T_c(X)] \rangle /\langle \tanh^{2}[J_{ij}/T_c(X)] \rangle ^{n}
\label{mu}
\end{equation}
It turns out that the effect of this deviation term is numerically rather weak in all dimensions.

If $d$ is high enough, $T_c$ values will also be high ( in $d = 8$ $T_c \sim 3.6$ in the examples discussed below) so the whole range of $[J_{ij}/T_c(X)]$ values can be taken to be small and we can evaluate $\tanh^2(x)$ for small arguments only. Now
\begin{equation}
\tanh^2(x)
= x^2 - (2/3)x^4 + (17/45)x^6 ...
\label{tanh}
\end{equation}
so keeping only the two leading terms, Eq.~(\ref{omegaX})  becomes 
\begin{equation}
\omega_c=\langle J_{ij}^2(X) \rangle /T_c(X)^2 - (2/3) \langle J_{ij}^{4}(X) \rangle /T_c^4
\label{omega}
\end{equation}
As all ISG interaction distributions are normalized such that $\langle J_{ij}^2(X) \rangle =1$ and in the high $d$ limit $\omega_c$ is independent of distribution, 
\begin{equation}
\omega_c = 1/T_c(X)^2 -(2/3)K(X)/(\omega_{c}T_c(X)^{4})
\end{equation}
with $K(X)$ the kurtosis of the distribution $X$. To the same level of approximation the equation can finally be rewritten
\begin{equation}
T_c(X,d)=(1/\omega_{c}(d))^{0.5}[1 - K(X)\omega_{c}(d)/3]
\label{approx}
\end{equation}
This should be exact in the limit of high $d$ and small to moderate kurtosis. The critical temperature should then vary
linearly with the kurtosis of the distribution $K(X)$, with a slope which is 
given by $[\omega_{c}(d)]^{0.5}/3$. $\omega_{c}(d)$ can be fixed if one $T_c(X)$ (generally the bimodal
value) is known.
\begin{figure}
\includegraphics[width=9cm, height=7cm,angle=0]{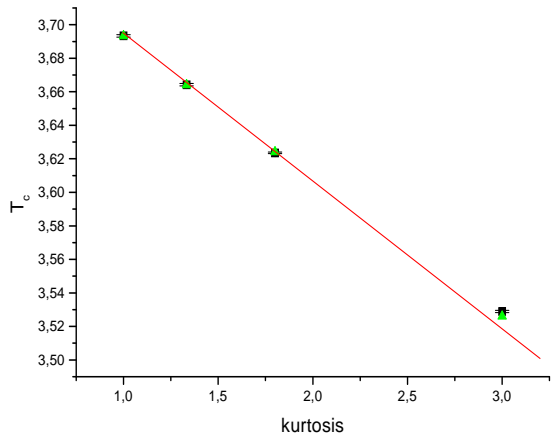}
\caption{(Colour on line) $T_c$ estimates for near neighbour interaction ISGs in dimension 8 from high temperature series data, Daboul {it et al} \cite{daboul:04} (black squares), the Singh-Fisher \cite{singh:88} estimates 
(see text) (green triangles), and  the leading approximation  Eq.~(\ref{approx}), straight line. The points for successive $K$ refer to ISGs 
with bimodal, double triangle, uniform and Gaussian distributions from left to right (see \cite{bernardi:97} for the definitions of the interaction distributions with the standard normalizations) } 
\protect\label{fig:D1}
\end{figure}

\begin{figure}
\includegraphics[width=9cm, height=7cm,angle=0]{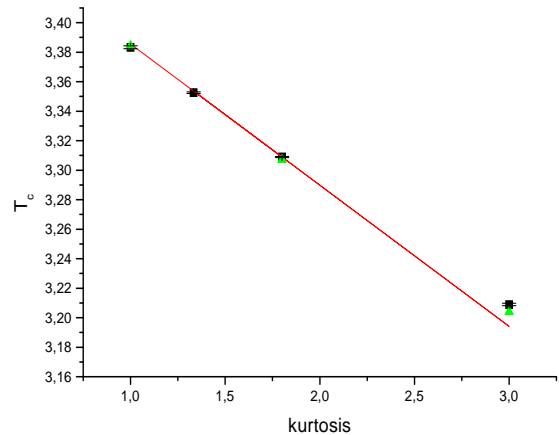}
\caption{(Colour on line) $T_c$ estimates for near neighbour interaction ISGs in dimension 7 from high temperature series data, Daboul {it et al} \cite{daboul:04} (black squares), the Singh-Fisher \cite{singh:88} estimates 
(see text) (green triangles), and  the leading approximation  Eq.~(\ref{approx}), straight line. The points refer to ISGs 
with bimodal, double triangle, uniform and Gaussian distributions from left to right.} 
\protect\label{fig:D2}
\end{figure}

I calibrate the single free parameter $\omega_{c}(d)$ by using the Daboul {\it et al} value for $T_c(\pm J)$ in each dimension (the Singh-Fisher $1/\sigma$ series $T_c(\pm J)$ values are in fact very close to the Daboul {\it et al} values), except in dimension 4 where the error bar on the Daboul {\it et al} value is high and I calibrate with the best simulation value. The $\omega_c$ values used are $0.069877, 0.08251, 0.1395$ and $0.2208$ in dimensions $8, 7, 5$ and $4$ respectively. The distributions studied by Daboul {\it et al} were the bimodal, double triangle, uniform and Gaussian with $K = 1, 4/3, 9/5$ and $3$ respectively. In simulations \cite{bernardi:97} the Laplacian (or decreasing exponential) distribution which has $K = 6$ was also studied. 

In Figures 1 and 2 I show the high temperature series $T_c$ values \cite{daboul:04} plotted
against kurtosis for dimensions $d = 8$ and $d = 7$. The leading approximation,  Eq.~(\ref{approx}), gives the straight line which lies very close to the Daboul {\it et al} data points for the different distributions. Using the full Singh-Fisher expressions (Eqs.~(\ref{omegaX}), ~(\ref{deviation}), and ~(\ref{mu})), evaluated by numerical integration without appealing to the small argument approximation gives calculated points which are in almost perfect agreement with the Daboul {\it et al} data in these dimensions. The weak deviation from the straight line for $K = 3$ (the Gaussian distribution) is due to the error introduced by truncating Eq.~(\ref{tanh}) in the wings of the Gaussian. 

By dimension 5, figure 3, the first order approximation has become much poorer as could be expected {\it a priori} because the small argument development is no longer valid in this lower dimension, where the $T_c$ values are lower. There are weak deviations of the Singh-Fisher estimates (without the small argument approximation) from the high temperature series data points,  meaning that Eq.~(\ref{deviation}) is no longer accurately obeyed, either because higher order terms give non-negligible contributions or more fundamentally because the $1/\sigma$ expansion becomes inaccurate at lower dimensions. As pointed out by Singh and Fisher their expansion "is at best asymptotic and optimum estimates come by truncating after the smallest term" 
\cite{singh:88}, a procedure which introduces uncertainties at lower dimensions. Nevertheless the general trend of decreasing $T_c$ with increasing kurtosis is still clearly present. 

\begin{figure}
\includegraphics[width=9cm, height=7cm,angle=0]{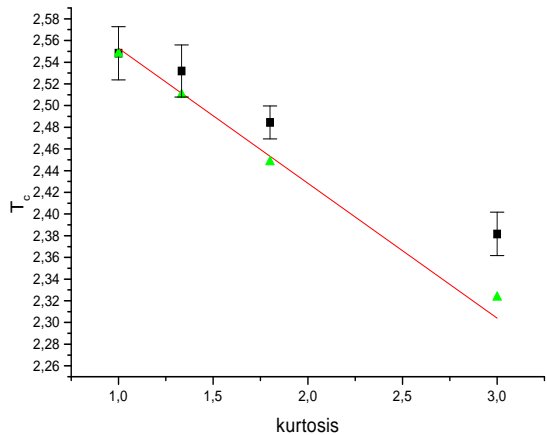}
\caption{(Colour on line) $T_c$ estimates for near neighbour interaction ISGs in dimension 5 from high temperature series data, Daboul {it et al} \cite{daboul:04} (black squares), the Singh-Fisher \cite{singh:88} estimates (see text) (green triangles), and  the leading approximation  Eq.~(\ref{approx}), straight line. The points refer to ISGs with bimodal, double triangle, uniform and Gaussian distributions from left to right.} 
\protect\label{fig:D3}
\end{figure}

Finally for $d=4$ the small argument approximation to the Singh-Fisher expressions is certainly inappropriate but one can again evaluate the full Singh-Fisher ordering temperature estimates through numerical integration. In this dimension ordering temperatures have also been estimated independently by simulations, using various complementary methods. These include a  technique relying on consistency between dynamic and static critical parameters \cite{bernardi:97},  domain wall analysis \cite{hukushima:99},  and Binder parameter measurements \cite{parisi:96,ney-nefle:98,marinari:99,young}. In the two cases (bimodal and Gaussian interaction distributions) where simulation results using different methods exist, agreement between the various simulation estimates is excellent, see Table I. In addition, the $T_c$ values have been studied in $d = 4$ (and $d = 3$) using a Migdal-Kadanoff (MK) approach \cite{prakash:97} where the optimum size $b^{*}$ of the MK renormalization cell is chosen by intrapolating the number of segments $b$ in an MK branch until $T_c(MK)$ is equal to $T_c$ for the $\pm J$ distribution. $T_c(MK)$ values for the other distributions are then estimated with this same $b^{*}$. For $d=4$ the optimum intrapolated value for $b$ was found to be $b^{*} = 2.49$ \cite{prakash:97} . The different $T_c$ estimates are given in Table 1 and are shown in Figure 4 where the same general trend once more appears. It can be seen that while the Singh-Fisher values and the MK values lie close to but not precisely on the series and simulation points, the agreement between the simulation estimates and the series values is excellent. As the simulation calculations are necessarily carried out on samples of moderate size (a maximum of $L=12$ up to now in dimension $4$ ), {\it a priori} there might have been a lingering suspicion that the $T_c$ estimates from simulations could somehow be intrinsically in error because of corrections to finite size scaling that had not been properly taken into account. Because the series estimates are not subject to this type of error, the agreement beween the two sets of quite independent $T_c$ values lays this suspicion to rest. One can have confidence that at least in dimension $d = 4$ the simulation estimates of $T_c(X)$  ( and in particular those using the static/dynamic consistency technique \cite{bernardi:97} which are the most complete ) are intrinsically correct and are not polluted by uncontroled corrections to finite size scaling. 

\begin{figure}
\includegraphics[width=9cm, height=7cm,angle=0]{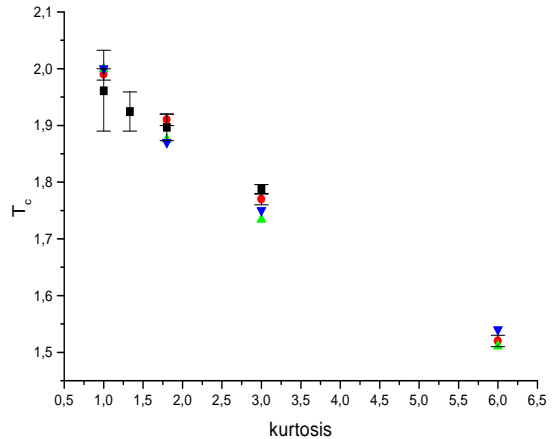}
\caption{(Colour on line) $T_c$ estimates for near neighbour interaction ISGs in dimension 4 from high temperature series data, Daboul {it et al} \cite{daboul:04} (black squares), the Singh-Fisher \cite{singh:88} estimates 
(see text) (green triangles), simulation estimates \cite{bernardi:97} (red circles), and MK estimates with $b^{*}=2.49$ \cite{prakash:97} (blue inverted triangles). The points refer to ISGs with bimodal, double triangle, uniform, Gaussian and Laplacian distributions from left to right.} 
\protect\label{fig:D4 }
\end{figure}

\begin{table}
\caption{\label{Table:1} Comparison of estimates of $T_c$ in dimension 4} 
\begin{tabular}{cccc}
distribution & method & reference & $T_c$\\ 
\hline
bimodal$(K=1)$ & series & \cite{klein:91} & 2.02(3) \\ 
bimodal & series & \cite{daboul:04} & 1.96(7) \\ 
bimodal & simulation & \cite{marinari:99} & 2.03(3)\\ 
bimodal & simulation & \cite{hukushima:99} & 2.00(4)\\ 
bimodal & simulation & \cite{young} & 2.00(1)\\ 
bimodal & simulation & \cite{bernardi:97} & 1.99(1)\\
triangles$(K=4/3)$ & series & \cite{daboul:04} & 1.95(1)\\
uniform$(K=9/5)$ & series & \cite{daboul:04} & 1.90(2)\\
uniform & simulation & \cite{bernardi:97} & 1.91(1)\\
uniform & MK & \cite{prakash:97} & 1.87(2) \\
Gaussian$(K=3)$ & series & \cite{daboul:04} & 1.78(1)\\ 
Gaussian & simulation & \cite{parisi:96} & 1.80(2)\\
Gaussian & simulation & \cite{ney-nefle:98} & 1.80(2)\\
Gaussian & simulation & \cite{bernardi:97} & 1.76(2)\\
Gaussian & MK & \cite{prakash:97} & 1.75(2)\\
Laplacian$(K=6)$ & simulation & \cite{bernardi:97} & 1.52(1)\\
Laplacian & MK & \cite{prakash:97} & 1.54(2)\\
\hline
\end{tabular}
\end{table}

Empirically, Table 1 shows that the MK $T_c$ values are a good indication of the true $T_c$ values if the appropriate choice of $b^{*}$ has been made by calibrating $b$ using the $T_c$ for one particular distribution (see the discussion in \cite{prakash:97}). In dimension $3$ where the series method would require a very long set of approximants, and where it turns out that in simulations corrections to finite size scaling are much stronger than in dimension $4$, it would appear that the ratios of $T_c$ values for different distributions obtained through the MK approach \cite{prakash:97,nogueira:99} should be rather reliable and a very helpful indication for controling simulation measurements. 

We can conclude that the key to understanding how $T_c$ values vary from distribution to distribution in ISGs with near neighbour interactions is the Singh-Fisher statement that $\omega_{c}(d)$ ( Eq.~(\ref{omegaX}) ) is almost distribution independent for fixed dimension $d$. The linear dependence of $T_c(X)$ on the kurtosis $K(X)$ with a slope $[\omega_{c}(d)]^{0.5}/3$ becomes exact in the high dimension limit and is still a close approximation in high but finite dimensions ($d=8$ and $7$). Even at moderate $d$ ($d=5$ and $4$), the general trend still holds so the ranking of distributions by their kurtosis remains physically appropriate in agreement with the initial phenomenological suggestion \cite{bernardi:97}.  ( It is nevertheless conceivable that for low dimensions  differences between distributions could lead to situations such that two distributions with identical kurtoses would have marginally different $T_c$ values. For the moment the number of different distributions that have been studied is very limited so this possibility has never been put to the test). In dimension $d = 4$ which is the only dimension where appropriate data is available for both simulations and series measurements, simulation estimates of ordering temperatures are in excellent agreement with the series values which confirms that the simulation estimates, in  particular those based on the consistency between dynamic and static critical parameters \cite{bernardi:97}, are accurate and reliable.

Acknowledgements : I would like to thank A. Aharony, D. Daboul, R.R.P. Singh and H.G. Katzgraber for their helpful comments.

\end{document}